\documentstyle[prl,aps,epsf]{revtex}
\begin{document}
\twocolumn[\hsize\textwidth\columnwidth\hsize\csname@twocolumnfalse%
\endcsname

\draft
\preprint{SU-ITP \# 97/33}
\title{Microscopic Electron Models with Exact SO(5) Symmetry} 

\author{Silvio Rabello, Hiroshi Kohno$^1$, 
Eugene Demler, and Shou-Cheng Zhang}
\address{\it Department of Physics, Stanford University, Stanford
  CA~~94305 \\ 
$^1$ Department of Physics, University of Tokyo, Bunkyo-ku, Tokyo 113, Japan} 
\maketitle
\begin{abstract}
\end{abstract}
{\sl We construct a class of microscopic electron models with
exact SO(5) symmetry between antiferromagnetic and $d$ wave 
superconducting ground states. There is an exact one-to-one correspondence 
between both single-particle and collective excitations in both phases.
SO(5) symmetry breaking terms can be
introduced and classified according to irreducible representations of
the exact SO(5) algebra. The resulting phase diagram and
collective modes are identical to that of the SO(5) nonlinear $\sigma$ model.}

\pacs{ PACS numbers: 74.20.Mn, 74.25.Ha, 71.10.-w }
]

One of the most interesting features of the high $T_c$ superconductivity
is the close proximity and the interplay between the antiferromagnetic
(AF) and the $d$ wave superconducting (dSC) phases. Recently, a theoretical
formalism was introduced based on the concept of a $SO(5)$ symmetry
between these two phases, and the resulting field-theoretical model
describes the cuprate phase diagram and collective modes in a unified
framework\cite{so5}. It was argued that the microscopic Hubbard model
supports an approximate $SO(5)$ symmetry\cite{so5,neutron,meixner}.

In this note, we construct a class of microscopic electron models with
exact $SO(5)$ symmetry. In this model, degeneracy between the AF and
dSC phases can be demonstrated exactly, and both the fermionic 
single-particle and the bosonic collective modes can be mapped onto
each other with a precise one-to-one correspondence. This model can be
used as a starting point around which $SO(5)$ symmetry breaking interactions
can be introduced and classified according to irreducible tensors of the
$SO(5)$ algebra. It is shown that the resulting phase diagram and the
collective modes are similar to those obtained from the effective $SO(5)$
nonlinear $\sigma$ model with anisotropic couplings\cite{so5,burgess}.
The purpose of this work is to demonstrate that the general $SO(5)$ idea can be
realized exactly by explicit microscopic Hamiltonians. The microscopic
information extracted from this class of models, especially the behavior
of the fermionic excitations across the AF/dSC transition would greatly 
complements the effective field theory approach. 
Within this class of models, we have
a consistent microscopic theory of the AF/dSC transition. Since 
both the AF and the dSC states are stable infrared fixed points, it is plausible
that one can deform the parameters so that the microscopic $SO(5)$
models can also serve as a paradigm for a much more general class of
AF/dSC transitions, including those occuring in the high $T_c$ 
cuprates and 2D organics.

The first independent attempt to construct microscopic $SO(5)$ 
invariant models was undertaken by Christopher Henley\cite{henley}.
He independently made a crucial observation that if one replaces 
the standard $d$ wave factor $\cos p_x - \cos p_y$ by
${\rm sgn}(\cos p_x - \cos p_y)$,
the $SO(5)$ algebra introduced in \cite{so5} closes exactly.
 
It is easy to write down many electron models with exact $SU(2)$ spin
rotation invariance, because the electron operator 
$c_{{\bf p} \sigma}$
forms a natural spinor representation of the $SU(2)$ algebra. In writing
down $SU(2)$ invariant models, all we have to do is to contract the spinor
indices in a natural way. Therefore, the first step towards constructing a
$SO(5)$ invariant electron model is to find a natural definition
of a $SO(5)$ spinor. Spinor representations of the $SO(5)$ Lie algebra
can be easily constructed using the Clifford algebra of five $4\times4$
Dirac matrices\cite{sp4} satisfying 
$ \{ \Gamma^a , \Gamma^b \}= 2\delta_{ab}~(a,b=1,...,5) $, 
and the ten $SO(5)$ rotation generators are given by 
$ \Gamma^{ab}=-i[\Gamma^a, \Gamma^b ] $. 
In this work we shall use the following explicit representation for the
Clifford algebra:
\begin{eqnarray}
 \Gamma^1\!=\! \left( \begin{array}{cc} 
               0          & -i\sigma_y  \\
               i\sigma_y  & 0            \end{array} \right)      
 \Gamma^{(2,3,4)} \!=\! \left( \begin{array}{cc} 
             \vec \sigma  & 0  \\
                  0       &  ^t\vec \sigma  \end{array} \right) 
 \Gamma^5 \!=\! \left( \begin{array}{cc} 
                0         & \sigma_y  \\
                \sigma_y  & 0            \end{array} \right)  
\nonumber
\end{eqnarray}
where $\vec \sigma = (\sigma_x, \sigma_y, \sigma_z)$ are the usual 
$2\times 2$ Pauli matrices, and ${}^t$ means transposition.  
We define a $4$ component spinor by 
\begin{eqnarray}
\label{spinor}
^t\Psi_{{\bf p}} = \left\{   
                  c_{{\bf p} \uparrow}   ,
                  c_{{\bf p} \downarrow} ,
        \phi_\pi ({\bf p}) c^{\dagger}_{{\bf -p+Q}, \uparrow}  ,
        \phi_\pi ({\bf p}) c^{\dagger}_{{\bf -p+Q}, \downarrow} 
                  \right\} 
\end{eqnarray}
where $\phi_\pi ({\bf p})={\rm sgn}(\cos p_x - \cos p_y) = \pm 1$, 
and ${\bf Q} = (\pi,\pi )$.
Since we have two spin 
degrees of freedom at a given momentum ${\bf p}$, such a 
description must be redundant. Indeed, one can easily see that the
spinors with momenta outside of the magnetic Brillouin zone is related to the
spinors inside the magnetic Brillouin zone by an ``R conjugation"
\begin{eqnarray}
\label{conjugation}
\Psi_{\bf p+Q} = \phi_\pi ({\bf p})R\,\Psi^*_{\bf -p} . 
\end{eqnarray} 
The $R$ matrix is an invariant tensor of the $SO(5)$ algebra enjoying the
following properties:
$
  R\,\Gamma^aR = -{}^t\Gamma^a, \ \ R\,\Gamma^{ab}R = {}^t\Gamma^{ab}
$. In our representation it takes the
form $R = \left( \begin{array}{cc} 
                     0  &  1  \\
                    -1  &  0  \end{array} \right)$.
(The existence of such a matrix is related to the fact that the spinor
representation of the $SO(5)$ Lie algebra is pseudo-real.
The $\sigma_y$ matrix plays a
similar role for $SO(3)$). The $\Psi_{{\bf p}\alpha}$ spinors obey the
anticommutation relations:
\begin{eqnarray}
\label{canonical}
\{ \Psi^{\dagger}_{{\bf p}\alpha}, \Psi_{{\bf p'}\beta} \}  
     = \delta_{\alpha \beta}\delta_{{\bf p,p'}} , \hskip 2cm    \\
\{ \Psi^{\dagger}_{{\bf p}\alpha}, \Psi^{\dagger}_{{\bf p'}\beta} \} = 
\{ \Psi_{{\bf p}\alpha}, \Psi_{{\bf p'}\beta} \}  
     = -\phi_{\pi}({\bf p}) R_{\alpha \beta} \delta_{{\bf p+p',Q}} .
\end{eqnarray}
If we restrict ${\bf p}$ and ${\bf p'}$ to be both inside the magnetic
Brillouin zone, 
the right hand side of the second equation vanishes and the
$\Psi_{{\bf p} \alpha}$ 
spinors commute in the same way as the $c_{{\bf p}\sigma}$ spinors. 
They can be used to construct the $SO(5)$ vector order parameter and
the symmetry generators:
$
\label{vector}
n_a = \frac{1}{4}\sum_{{\bf p}} w_{\bf p} \Psi^{\dagger}_{{\bf p}} \Gamma^a
                                \Psi_{{\bf p+Q}}
$, and $ \label{generator}
L_{ab} = \frac{1}{8}\sum_{{\bf p}} \Psi^{\dagger}_{{\bf p}} \Gamma^{ab}
                                   \Psi_{{\bf p}}. 
$
Here $w_{\bf p}=w_{-\bf p}$.
Note that the definition of the $\pi$ operators
  ($L_{1\,(2,3,4)},L_{(2,3,4)\,5}$)
differs from the ones used in previous works\cite{so5,neutron,meixner}, 
where they are electron pair operators on the nearest neighbor (n.n.) sites. 
 The problem with this kind of definition is that the commutator algebra 
does not close, and generates longer-ranged bonds. Naively, the condition for 
the closure of the $SO(5)$ algebra appears to be over-constrained. 
 The present work and \cite{henley} start with electron pair operators with a long-ranged
profile, given in real space by the lattice Fourier transform of 
$\phi_\pi ({\bf p})$, 
\begin{eqnarray}
  \phi_\pi (m,n) 
                 = \frac{2}{\pi^2}\frac{1-(-)^{m+n}}{m^2-n^2},   
\end{eqnarray}
where ${\bf R}=(m,n)$ is a lattice point. 
It is truly remarkable that this simple choice closes the algebra exactly.
Notice that while the $\pi$ operators have long-ranged profiles, the dSC 
order parameter can still be short-ranged with suitable choices of $w_{\bf p}$. 
Under the $SO(5)$ rotations generated by the $L_{ab}$, $\Psi_{{\bf p}}$ 
transforms as a proper $SO(5)$ spinor
\begin{eqnarray}
 [ L_{ab}, \Psi_{{\bf p} \alpha}] = 
 - \frac{1}{4}(\Gamma^{ab})_{\alpha\beta} \Psi_{{\bf p} \beta}
\end{eqnarray}
for all values of ${\bf p}$.
Using these spinors, exact $SO(5)$ invariant Hamiltonians can be constructed
simply by proper contraction of the spinor indices.

We start with the kinetic term, and write it as 
\begin{eqnarray}
\label{kinetic}
 H_{{\rm kin}} = \sum_{{\bf p},\sigma} \varepsilon_{{\bf p}} 
                c^{\dagger}_{{\bf p},\sigma} c_{{\bf p},\sigma}
               = \frac{1}{2}\sum_{{\bf p}} \varepsilon_{{\bf p}} 
                \Psi^{\dagger}_{{\bf p}} \Psi_{{\bf p}}. 
\end{eqnarray}
We see that the property 
$\varepsilon_{{\bf p+Q}}=-\varepsilon_{{\bf p}}$,
valid for n.n. tight binding model, 
is crucial for this construction to work. 
In order to construct four-fermion interactions, we first 
note that a $SO(5)$ spinor bilinear can in general be decomposed into
a direct sum of a scalar,  a vector, and 
an antisymmetric tensor, {\it i.e.} $4\times 4=1+5+10$.
Therefore, general $SO(5)$ invariant four-fermion interactions can be
expressed as 
\begin{eqnarray}
\label{interaction}
 H_{{\rm int}} = &{}&\sum_{{\bf p, p', q}} V_1({\bf p},{\bf p'};{\bf q}) 
       (\Psi^{\dagger}_{{\bf p}} \Gamma^{a}\Psi_{{\bf p+q}})
       (\Psi^{\dagger}_{{\bf p'}} \Gamma^{a}\Psi_{{\bf p'-q}}) \nonumber\\
  &+& \sum_{{\bf p, p', q}} V_2({\bf p},{\bf p'};{\bf q}) 
       (\Psi^{\dagger}_{{\bf p}} \Gamma^{ab}\Psi_{{\bf p+q}})
       (\Psi^{\dagger}_{{\bf p'}} \Gamma^{ab}\Psi_{{\bf p'-q}}) \nonumber\\
  &+& \sum_{{\bf p, p', q}} V_0({\bf p},{\bf p'};{\bf q})
       (\Psi^{\dagger}_{{\bf p}}  \Psi_{{\bf p+q}})
       (\Psi^{\dagger}_{{\bf p'}} \Psi_{{\bf p'-q}}). 
\label{sym}
\end{eqnarray}
Since $
 L_{ab}({\bf p},{\bf q}) \equiv \Psi_{{\bf p}}^{\dagger} \Gamma^{ab}
   \Psi_{{\bf p}+{\bf q}}$,
 $n_a({\bf p},{\bf q}) \equiv  \Psi_{{\bf p}}^{\dagger} \Gamma^{a}
   \Psi_{{\bf p}+{\bf Q}+{\bf q}}$ and $
 \rho ({\bf p},{\bf q}) \equiv  \Psi_{{\bf p}}^{\dagger} \Psi_{{\bf p}+{\bf q}}
$
 are the true $SO(5)$ tensor, vector, and scalar, respectively, for any
${\bf p}$ and ${\bf q}$, their inner products naturally gives a manifestly
$SO(5)$ invariant Hamiltonian.

Among three terms in $H_{{\rm int}}$, 
we concentrate on the vector interaction (first term) 
in all subsequent analysis, and assume a factorizable form 
$V_1({\bf p}, {\bf p'}; {\bf q}) = - V_1({\bf q}) w_{{\bf p}} w_{{\bf p'}}$;
This form is not necessary, but simplifies calculations. 
In real space, 

\begin{eqnarray}
\label{H1R}
  H_{{\rm int},1}  = - 4\sum_{\ell,n} V_1({\bf R}_\ell - {\bf R}_n) 
                 {\rm e}^{i{\bf Q} \cdot ({\bf R}_\ell - {\bf R}_n)}
\nonumber\\
     \ \ \times \biggl[ \vec m_\ell \cdot \vec m_n + 
     \frac{1}{2} (\Delta_\ell \Delta^{\dagger}_n 
                 + \Delta^{\dagger}_\ell \Delta_n) \biggr] . 
\end{eqnarray}
 Here, $\vec m_\ell$ and $\Delta_\ell$ are Neel and $d$-wave 
pairing order parameters (operators) at site 
$\ell \equiv {\bf R}_\ell$, but with extended internal structures 
determined by $w_{{\bf p}}$. 
 For the simplest choice $w_{{\bf p}} = 1$, they become 

\begin{eqnarray}
\label{w1}
  \vec m_{\ell} &=& \frac{1}{2}
      ( \psi^{\dagger}_{\ell} \vec \sigma \psi_{\ell}
      - \chi^{\dagger}_{\ell} \vec \sigma \chi_{\ell} ) 
       {\rm e}^{i{\bf Q} \cdot {\bf R}_\ell},  \\
  \Delta^{\dagger}_{\ell} 
  &=& \sum_j \phi_\pi ({\bf R}_{\ell} - {\bf R}_j)
      ( c^{\dagger}_{\ell \uparrow} c^{\dagger}_{j \downarrow}
      - c^{\dagger}_{\ell \downarrow} c^{\dagger}_{j \uparrow}).             
\end{eqnarray}
 Here we introduced two-component spinors 
$\psi_\ell = {}^t (c_{\ell \uparrow} , c_{\ell \downarrow})$ 
and 
$\chi_\ell = (-{\rm e}^{i{\bf Q} \cdot {\bf R}_\ell}) \times  
 {}^t (b_{\ell \uparrow} , b_{\ell \downarrow})$
with $ b_{\ell \sigma} = 
 \sum_j \phi_{\pi}({\bf R}_\ell - {\bf R}_j) c_{j \sigma}$.
The pair wave function for dSC condensate is described by $\phi_\pi$
and is long-ranged. 
For the choice
$w_{{\bf p}} = |\cos p_x - \cos p_y|$, we obtain  

\begin{eqnarray}
\label{w2}
  \vec m_{\ell} &=& \frac{ {\rm e}^{i{\bf Q} \cdot {\bf R}} }{2}
      \sum_i \phi_M ({\bf R}_{\ell} - {\bf R}_i)
      ( \psi^{\dagger}_i \vec \sigma \psi_{\ell}
      - \chi^{\dagger}_i \vec \sigma \chi_{\ell} )  \\
  \Delta^{\dagger}_{\ell} 
  &=& \sum_{i,j} \phi_M ({\bf R}_{\ell} - {\bf R}_i)
                 \phi_\pi ({\bf R}_{\ell} - {\bf R}_j)
       ( c^{\dagger}_{i \uparrow} c^{\dagger}_{j \downarrow}
       - c^{\dagger}_{i \downarrow} c^{\dagger}_{j \uparrow})             
\end{eqnarray}
where
\begin{eqnarray}
\label{SDW}
\phi_M (m,n)
                 = \frac{4}{\pi^2}\!
                  \frac{1+(-)^{m+n}}{ \bigl[ (m+n)^2-1 \bigr] 
                                      \bigl[ (m-n)^2-1 \bigr] } . 
\label{fi-m}
\end{eqnarray} 
The interaction between centers of mass of $\vec m$ or $\Delta$ 
fields is controlled by $V_1({\bf R})$. 
 If we take $V_1({\bf q})$ to be a $\delta$ function at ${\bf q} = {\bf Q}$, 
the $\Delta$-part in $H_{{\rm int},1}$ becomes 
the usual BCS reduced Hamiltonian for n.n.~$d$-wave pairing.  
 If $V_1({\bf q})$ is taken to be a Lorenzian around ${\bf q} = {\bf Q}$,
the real space form of the spin interaction resembles the potential induced
by the AF paramagnon exchange\cite{scalapino,pines,swz}.

It is not difficult to find degeneracy between AF and dSC states
in the usual treatment of mean field theories. However, their excitation
spectra are generally different, and quantum fluctuations may remove this
degeneracy. In the $SO(5)$ invariant models, symmetry not only ensures
exact degeneracy of the ground states, but also ensures exact one-to-one
correspondence between their excitation spectra. This fact is formulated
as follows:

{\it Theorem 1:} If $|\Psi_0>$ is a ground state of a $SO(5)$ invariant
Hamiltonian with AF broken symmetry (say in the $n_2$ direction), 
{\it i.e.} $<\Psi_0|n_a|\Psi_0>=\delta_{2,a}A$, then 
$|\Psi_0'>=e^{i\frac{\pi}{2}L_{12}}|\Psi_0>$ is a degenerate ground
state with dSC broken symmetry (in $n_1$ direction), {\it i.e.} 
$<\Psi_0'|n_a|\Psi_0'>=\delta_{1,a}A$. Furthermore, all excited states
of the AF ground state can be mapped to excited states of the dSC
ground state at the same energy by the $e^{i\frac{\pi}{2}L_{12}}$
operator.

The proof of this theorem is elementary, since $L_{12}$ commutes with
the Hamiltonian, and $e^{-i\frac{\pi}{2}L_{12}} n_1 e^{i\frac{\pi}{2}L_{12}}
=n_2$. In the following, we shall illustrate this powerful theorem in an 
explicit mean field calculation.
We take a ``generalized BCS reduced Hamiltonian" by selecting 
$ V_1({\bf q}) =  V_1 \delta_{\bf q, \bf Q} $ in the vector interaction. 
The Green's function in the presence of a mean field 
$\langle n_{\bf p}^a\rangle=\frac{1}{4}\langle\Psi_{\bf p}^\dagger
\Gamma^a\Psi_{\bf p+Q}\rangle$ is given by

\begin{eqnarray}
  \label{Gfunc}
  G_{\alpha\beta}({\bf p},{\bf p}';\omega)&=&-i\int\, dt e^{
    i\omega t}
\langle T\Psi_{{\bf p},\alpha}(t)\Psi_{{\bf p}',\beta}^\dagger(0)\rangle
\\ &=&\frac{(\omega +\varepsilon_{\bf p})
\delta_{\alpha\beta}\delta_{{\bf p},{\bf p}'}+\Delta_{\bf p}^a
\Gamma^a_{\alpha\beta}\delta_{{\bf p},{\bf p}'+{\bf Q}}}
{\omega^2-\varepsilon_{\bf p}^2-{\Delta_{\bf p}^a}^2+i\delta}\, 
\nonumber
\end{eqnarray}
where $\Delta^a_{\bf p}= -16 V_1 w_{\bf p}\sum_{\bf k}w_{\bf k}
\langle n^a_{\bf k}\rangle$. This manifestly $SO(5)$ invariant 
Green's function explicitly shows that the AF quasi-particles can be
mapped onto dSC quasi-particles. In particular, the AF Green's function
(in the $n_2$ direction) can be obtained directly from the dSC Green's
function (in the $n_1$ direction) by a simple 
rotation: $G^{AF}= 
e^{-i\frac{\pi}{2}\Gamma_{12}} G^{SC} e^{i\frac{\pi}{2}\Gamma_{12}}$.
If we take $w_{\bf p}=1$, the AF quasi-particles have a full $s$ wave
gap, while the dSC quasi-particles have a full $d$ wave gap, with step
discontinuity at $(\pm \pi/2, \pm \pi/2)$ points. For the choice of
 $w_{\bf p}=\vert \cos p_x-\cos p_y\vert$, the dSC quasi-particles
have the usual  $\cos p_x-\cos p_y$ gap behavior (Fig.1), while the AF
quasi-particles have an anisotropic $s$ wave gap with nodes at 
$(\pm \pi/2, \pm \pi/2)$ points (Fig.2). Because the AF nodes are not 
``topological", any small interactions 
will
remove it\cite{coleman}.
In either case, the amplitude of the gaps
are the same in both phases.

In addition to the gapped single-particle excitations, 
the $SO(5)$ invariant models also have
four gapless Goldstone modes. There are ten conserved
Noether current density $L^{ab}_\mu(x)$  associated
with the $SO(5)$ symmetry, where  $\mu \, (=0,1,2)$ is the space-time index. 
In the small $q$ limit, exact current conservation leads to 
the following generalized Ward identity for the vertex function 
$\gamma^{ab}_\mu(p+q,p)$ ($p=(\omega,{\bf p})$) related to this current
\cite{schrieffer}
 
\begin{eqnarray}
  \label{Ward}
  q^\mu\biggl(\gamma^{ab}_\mu(p+q,p)\biggr)_{\alpha\beta}\!=\!
-\frac{1}{4}\biggl( 
\Gamma^{ab}_{\alpha\nu}G^{-1}_{\nu\beta}(p+q)-G^{-1}_{\alpha\nu}(p)
\Gamma^{ab}_{\nu\beta}\biggr)\, \nonumber
\end{eqnarray}
For a non-zero $\langle n^a_{\bf k}\rangle$, the rhs of the above equation
has a finite $q\rightarrow 0$ limit, therefore, the four vertex functions
$\gamma^{ab}_\mu(p+q,p)$ with $b\ne a$ must have gapless poles
in this limit. This result is formulated in the following theorem:   

{\it Theorem 2:} In a $SO(5)$ invariant model, there are four
gapless Goldstone modes associated with spontaneous symmetry breaking.

In the AF phase, there are two spin wave modes and 
a $\pi$ doublet ($4=2+2$). In the dSC phase, there is one SC phase mode,
and a $\pi$ triplet ($4=1+3$)\cite{so5}.

 As symmetry-breaking perturbations to the above $SO(5)$ invariant Hamiltonian, 
we consider two typical terms. 
 One is the coupling to external fields $B_{ab}$, 
\begin{eqnarray}
  H_{ext} = - \sum_{a<b} B_{ab} L_{ab}. 
\label{ext}
\end{eqnarray}
A particular example of this field is the chemical potential 
$B_{15}= - 2 \mu$, leading to $H_{\mu} = 2 \mu L_{15}$.
 The other is the anisotropy energy 
\begin{eqnarray}
\label{g-term}
  H_g  = - \sum_{{\bf p, p',q}}\!\sum_{a=2,3,4}\! g({\bf q})  
       (\Psi^{\dagger}_{{\bf p}}\! \Gamma^{a}\Psi_{{\bf p+q}})
       (\Psi^{\dagger}_{{\bf p'}}\! \Gamma^{a} \Psi_{{\bf p'-q}})   
\label{H_g}
\end{eqnarray}
between AF and dSC states. 
To study the spectrum of  
$ H = H_{{\rm kin}} + H_{{\rm int,1}} + H_\mu + H_g $, 
we take $g({\bf q}) = g\delta_{{\bf q},{\bf Q}}$ and again use mean field approximation. 
In the dSC phase, $\langle n_{\bf p}^a \rangle$ lies in the plane $(n_1,n_5)$. 
We choose it in the $n_1$ direction. Then the Green's function is given by  
\begin{eqnarray}
\label{Gsc}
   G^{SC}({\bf p},{\bf p'},\omega) \!=\!
               \left( \begin{array}{cc} 
               \frac{(\omega+\varepsilon_{\bf p}-\mu)1\delta_{{\bf p},{\bf p}'}}
                    {\omega^2 - (\varepsilon_{\bf p}-\mu)^2 - \Delta_{\bf p}^2 
                      + i\delta} &
 \frac{-i\sigma_y\Delta_{\bf p}\delta_{{\bf p},{p'+Q}}}{\omega^2-
(\varepsilon_{\bf p}+\mu)^2-\Delta_
{\bf p}^2+i\delta}\\ \frac{i\sigma_y\Delta_{\bf p}
\delta_{{\bf p},{\bf p}'+{\bf Q}}}{\omega^2-
(\varepsilon_{\bf p}-\mu)^2-\Delta_{\bf p}^2+i\delta}
& \frac{(\omega+\varepsilon_{\bf p}+\mu)1\delta_{{\bf p},{\bf p}'}}{\omega^2-
(\varepsilon_{\bf p}+\mu)^2-\Delta_
{\bf p}^2+i\delta}
\end{array} \right)\, \nonumber 
\end{eqnarray} 
where  
$\Delta_{\bf p}= -16 V_1\phi_\pi({\bf p}) w_{\bf p} 
 \sum_{\bf k}w_{\bf k} \langle n^1_{\bf k}\rangle 
 \equiv \Delta_0 \phi_\pi({\bf p}) w_{\bf p}$ 
and the $g$ term drops out because of symmetry mismatch. 
 $\Delta_0$ is determined by the gap equation 
$1 = 16 V_1\sum_{\bf k}\frac{w^2_{\bf k}}{2 E_{\bf k} }$, 
where $E_{\bf k} = \sqrt{(\varepsilon_{\bf k}-\mu)^2 + \Delta_{\bf k}^2 }$.  
For the choice $w_{\bf p}=\vert \cos p_x-\cos p_y\vert$, we have a 
usual $d$-wave gap. 
In the AF phase,   
$\langle n_{\bf p}^a\rangle$ lies in the $(n_2,n_3,n_4)$ space. 
If we pick the $n_4$ direction, we have 
\begin{eqnarray}
  \label{Gaf}
&&G^{AF}({\bf p},{\bf p}',\omega)= \nonumber\\
&&\left(\! \begin{array}{cc} 
               \frac{(\omega_+ + \varepsilon_{\bf p})1\delta_
{{\bf p},{\bf p}'}+
\Delta_{\bf p}\sigma_z\delta_
{{\bf p},{\bf p}'+{\bf Q}}}{\omega_+^2-\varepsilon_{\bf p}^2-\Delta_
{\bf p}^2+i\delta} & 0 \\
                 0        &                         
               \frac{(\omega_- + \varepsilon_{\bf p})1\delta_
{{\bf p},{\bf p}'}+
\Delta_{\bf p}\sigma_z\delta_
{{\bf p},{\bf p}'+{\bf Q}}}{\omega_-^2-\varepsilon_{\bf p}^2-\Delta_
{\bf p}^2+i\delta}\end{array}\! \right)\ \nonumber  
\end{eqnarray}
Here $\omega_{\pm} = \omega \pm \mu $ and  
$\Delta_{\bf p}= -16 \sum_{\bf k}(V_1 w_{\bf p}w_{\bf k}+g) 
  \langle n^4_{\bf k}\rangle 
  \equiv w_{\bf p}\Delta_0+\Delta_g$. 
$\Delta_0$ and $\Delta_g$ are determined by the gap equation 
$ \Delta_{\bf p} = 16 \sum_{\bf k}
  (V_1 w_{\bf p} w_{\bf k} + g) \frac{\Delta_{\bf k}}{2E_{\bf k}}$. 
The real space form of $\Delta_{\bf p}$ has an on-site contribution from
$\Delta_g$ and a long-ranged contribution proportional to
$\Delta_0  \phi_M ({\bf R})$. We see that the $g$ terms leave dSC gap
unaffected, while it removes the AF gap node (See Figs.1 and 2).
The ground state energy curves are shown in Fig.3. 
The ``superspin flop" transition from AF to dSC occurs at $\mu=0$
for $H_g=0$, while it occurs at a finite value of $\mu_c$ for
$g\ne 0$. In this case, the AF/dSC transition is first order, with a finite
jump in hole density $x_c$ (See Figs.3 and 4).

While the above pictures are based on the mean field 
approximation, some exact statements
can be made about the AF/dSC transitions. $SO(5)$ is a rank 2 algebra,
and we can choose $Q=-L_{15}$ and $S_z=-L_{23}$ as the members of the 
Cartan (maximal commutative) subalgebra. 
In addition, we
have the Casimir operator $C=\sum_{a<b} L_{ab}^2$, which commutes
with all the generators and has eigenvalue $l(l+3)$. 
The set $(Q,S_z,C)$ forms a Cartesian coordinate system  
labelling the quantum numbers of all states in the Hilbert space. 
If we consider only states with
even number of electrons, these states form a pyramid, with the $l=0$
singlet on top, the $l=1$ vector next, and the $l=2$ traceless
symmetric tensor on the 3rd layer etc. States on the same layer are all
connected by repeated actions of the 8 root generators. Therefore,

{\it Theorem 3:} In $SO(5)$ invariant models, it is sufficient to diagonalize
the Hamiltonian at half-filling with $Q=0$ and $S_z=0$. All the other states
(with even number of electrons) in the Hilbert space can be obtained  
from these states through the action of $(S_x,S_y,\vec \pi, \vec \pi^{\dagger})$.

In this sense, states at half-filling fully determine the states away from
half-filling. In the presence of the $H_\mu$ term, 
the $\pi^\dagger_\alpha$ and $\pi_\alpha$ operators are exact eigenoperators 
of the Hamiltonian with eigenvalue $\pm 2\mu$. Therefore, 
$H_\mu$ commutes with the Casimir operator, and
simply shifts the energy of the 
$Q\ne 0$ states linearly without changing the wave function of these states. 
In a system with spontaneous symmetry breaking, lowest states with different 
$l$ quantum numbers are separated by inverse system size. In the infinite
system limit, these shifts due to chemical potential converge to the
parabola as depicted in \mbox{Fig. 3\cite{bonner}}.

The symmetry-breaking terms, $H_g$ and $H_\mu$ produce the mass gap in the 
Goldstone mode spectrum. For $H_g=0$, the mass of the $\pi$ triplet
mode is exactly $2 |\mu|$. For finite $H_g$ we employ the
equations of motion (EOM).

As discussed earlier, $L_{ab}$'s  commute with $H_{{\rm kin}} + H_{{\rm int, 1}}$ 
for arbitrary 
$V_1 ({\bf p},{\bf p}';{\bf q})$. 
The only contributions to the EOM for $L_{ab}$ come from the 
$SO(5)$ symmetry breaking terms, $H_g$ and $H_{\mu}$:  
\begin{eqnarray}
\label{L-eq}
[ L_{ab}, H ]& = &- i \sum_c ( B_{ac} L_{bc} - B_{bc} L_{ac} )  \\
&-& i ( \delta_a - \delta_b ) \sum_{{\bf q}} g({\bf q+Q})
    \{ n_a ({\bf q}),  n_b (-{\bf q}) \} ,  
      \nonumber
\end{eqnarray}
where $\delta_a =1$ for $a=2,3,4$, and $=0$ otherwise. 
This is precisely the same equation as that obtained in \cite{so5}. 
EOM for $n_a$ is given by 
\begin{eqnarray}
\label{n-eq}
&{}&[ n_{a}, H ] \!=\! \frac{1}{4}\! \sum_{\bf p}\! 
   ( \varepsilon_{{\bf p}+{\bf Q}} - \varepsilon_{\bf p} ) 
   n_a ({\bf p},0) \!+\!  i \sum_b B_{ab} n_b   \\
 &-&\! i\! \sum_{ {\bf q}, b}\! \left( V_1({\bf q+Q}) +
   g({\bf q+Q})  \delta_b \right)  
     \{ n_b ({\bf q}),   L_{ab}(-{\bf q})\} . \nonumber
\end{eqnarray}
where 
$ n_a({\bf q}) = \frac{1}{4} \sum_{\bf p}\! n_a({\bf p},{\bf q}) ,  
  L_{ab}({\bf q}) = \frac{1}{8} \sum_{\bf p} L_{ab}({\bf p},{\bf q}) $, 
and a form $V_1({\bf p},{\bf p}';{\bf q})= - V_1({\bf q})$ 
and the property $V_1({\bf Q}-{\bf q}) = V_1({\bf Q}+{\bf q}) $ has been assumed.

Let us now consider the case where only the chemical potential 
$B_{15} = - 2 \mu$ is introduced as external fields, 
and consider only the collective part in equation (\ref{n-eq})\cite{Anderson58,eta}.
In the dSC state, we take
$ \langle n_a ({\bf q}) \rangle = \langle n_1 \rangle 
                                  \delta_{a,1} \delta_{{\bf q,0}}$, 
and linearize eq.(\ref{n-eq}) to obtain 
\begin{eqnarray}
 \dot{n}_{a} &=& 2 V_1 \langle n_1 \rangle  L_{1a} , \hspace{1cm}
 (a=2,3,4) \\ 
\label{na}
 \dot{n}_{5} &-& 2 \mu =  2 V_1  \langle n_1 \rangle L_{15} ,
\label{n5}
\end{eqnarray}
where $V_1 \equiv V_1({\bf Q})$. 
These equations should be compared with  
$ \chi \dot{n}_{a} =   L_{1a}$ with $a=2,3,4 $ and $ \chi ( \dot{n}_{5}
 - 2 \mu ) =  L_{15} $ derived from the nonlinear $\sigma$-model\cite{so5}. 
 Equations (\ref{L-eq}), (\ref{na}), and (\ref{n5}) can be combined to
give
$\ddot{n}_5 = 0$
and
$\ddot{n}_a = 4 (\mu_c^2 - \mu^2) n_a $, where 
$ \mu_c = \langle n_1 \rangle \sqrt{ g V_1}$ and $g \equiv g({\bf Q})$.
Therefore, the energy of the triplet $\pi$ excitations in the dSC
state is given by 
\mbox{$\omega_0 = 2\sqrt{ \mu^2 - \mu_c^2 }$}, which is also consistent with 
the result of \cite{so5}. 
Similar calculation in the AF state gives the energies of the $\pi$ doublet
\mbox{$
\omega_0 = 2 \langle n_4 \rangle \sqrt{ g \left(  V_1 + 
  g \right)} \pm 2 \mu
$}, where we assumed AF ordering along $n_4$.
We therefore see that the two symmetry breaking terms $g$ and $\mu$ partially
compensate each other for the $\pi$ triplet and $Q=-2$ $\pi$ doublet. 

In conclusion we have constructed a class of electron models with exact
$SO(5)$ symmetry. Both the fermionic single particle and bosonic
collective modes of the AF and dSC phases are in one-to-one 
correspondence to each other. Energy levels are classified according to the
$SO(5)$ quantum numbers and the level crossing at the AF/dSC transition
can be followed in detail. The fermionic single-particle gaps do not close
at the AF/dSC transition, but instead rotate from one direction in superspin
space  to another. It is amusing to ask what experimental results this
kind of ideal models would predict. The phase diagram would be identical
to the one depicted in Fig. 1A of reference \cite{so5}, with an insulating AF
gap and a finite jump in chemical potential at half filling, phase separation
or stripe ordering for doping range $0<x<x_c$, a low energy spin triplet
$\pi$ resonance in the dSC phase, and a ``pseudogap" behavior in the high 
temperature phase above the bi-critical point, where the gap direction
fluctuates (rotates) between AF and dSC character.

The authors are deeply indebted to Prof. C. Henley for generous sharing of his
ideas. We would like to thank Prof. R.B. Laughlin for focusing our attention
on the study of the AF/dSC transition problem, 
Prof. I. Affleck, E. Fradkin, D. Gross,  M. Oshikawa, J. Preskill,
A. Zee for useful discussions on $SO(5)$ Lie algebra and Prof. J. Berlinsky,
H. Fukuyama, C. Kallin, S. Kivelson and D. Scalapino for general 
discussion on the high-Tc problem. 
This work is supported by the NSF under grant numbers DMR-9400372 
and DMR-9522915. S.R. acknowledges the support from CNPq
(Brazilian Research Council).

\hspace{1cm}

\begin{figure*}[h]
\centerline{\epsfysize=3.5cm
\epsfbox{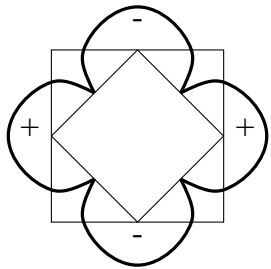}
}
\nonumber
\caption{ The superconducting (dSC) gap. }
\end{figure*}
\begin{figure}[h]
\centerline{\epsfysize=3.5cm
\epsfbox{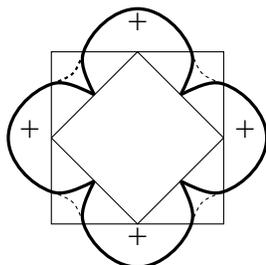}
}
\caption{The antiferromagnetic (AF) gap. The solid (dotted) line is for the case with 
(without) $SO(5)$ symmetry. }
\end{figure}
\begin{figure}[h]
\centerline{\epsfysize=3.5cm
\epsfbox{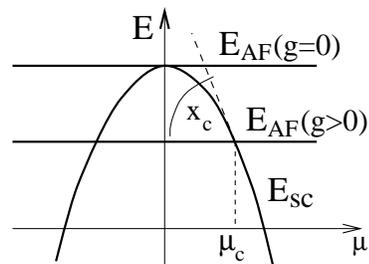}
}
\caption{ The groundstate energy $E_G$ in both AF and SC phases as functions of
$\mu$. }
\end{figure}
\begin{figure}[h]
\centerline{\epsfysize=3.5cm
\epsfbox{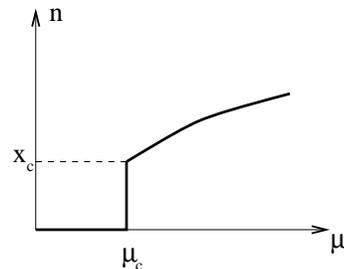}
}
\caption{The electron density $x$ versus $\mu$ in the presence of  
anisotropy energy $H_g$.}
\end{figure}


\begin{thebibliography}{10}

\bibitem{so5} S.C. Zhang,
{\em Science}, 275:1089, 1997.

\bibitem{neutron} E.~Demler and S.C. Zhang,
{\em Phys. Rev. Lett.}, 75:4126, 1995.

\bibitem{meixner} S.~Meixner, W.~Hanke, E.~Demler and S.C. Zhang,
preprint, cond-mat/9701217.

\bibitem{burgess} C.P. Burgess and C.A.Lutken,
preprint, cond-mat/9705216

\bibitem{henley} C. Henley, unpublished lecture notes, Cornell university.

\bibitem{sp4} Alternatively, one can also work out the spinor representation
from $Sp(4)$, the universal covering group of $SO(5)$.

\bibitem{scalapino} D. Scalapino, {\em Phys. Rep.}, 250:329, 1995;

\bibitem{pines} D. Pines, {\em Physica},  C235:113, 1994; 

\bibitem{swz} J. R. Schrieffer, X.G. Wen and S.C. Zhang,
{\em Phys. Rev.} B39:11663, 1989.

\bibitem{schrieffer} J.R. Schrieffer, in {\em Theory of Superconductivity},
( Benjamin, Reading, MA, 1964 ) 

\bibitem{coleman} One of us (SCZ) would like to thank P. Coleman,
F.D.M. Haldane and G. Murthy for a stimulating discussion on this
point.

\bibitem{bonner} J.C. Bonner and M.E. Fisher, 
{\em Phys. Rev.} 135:A640, 1964.

\bibitem{Anderson58} P.W. Anderson,
{\em Phys. Rev. }, 112:1900, 1958.

\bibitem{eta} E.~Demler, S.C. Zhang, N.Bulut, and D.J. Scalapino,
{\em Int. Journal of Modern Physics B}, 10:2137, 1996


\bibitem{com} It is possible to apply this kind of analysis to the $t$-$J$ 
Hamiltonian, and classify its operator content according to the $SO(5)$ algebra.  
This will be reported elsewhere. 

\end{thebibliography}
\end{document}